\documentclass[conference]{IEEEtran}
\IEEEoverridecommandlockouts
\usepackage{soul}
\usepackage{cite}
\usepackage{amsmath,amssymb,amsfonts}
\usepackage{algorithmic}
\usepackage{graphicx}
\usepackage{caption}
\usepackage{subcaption}
\usepackage{textcomp}
\usepackage{booktabs} %

\usepackage{xcolor}
\usepackage{url}
\usepackage{todonotes}
\usepackage{epstopdf}

\usepackage{listings}
\usepackage{courier}
\usepackage{color}
\definecolor{codegreen}{rgb}{0,0.5,0}
\definecolor{codegray}{rgb}{0.5,0.5,0.5}
\definecolor{codepurple}{rgb}{0.58,0,0.82}
\definecolor{backcolour}{rgb}{0.95,0.95,0.92}
\lstdefinestyle{mystyle}{
	basicstyle={\ttfamily\footnotesize},
	commentstyle=\color{codegreen},
	keywordstyle=\color{magenta},
	numberstyle=\tiny\color{codegray},
	stringstyle=\color{codepurple},
	rulecolor=\color{blue},
	breakatwhitespace=false,
	breaklines=true,
	captionpos=b,
	keepspaces=true,
	numbers=left,
	numbersep=5pt,
	showspaces=false,
	showstringspaces=false,
	showtabs=false,
	tabsize=2
}
\def\BibTeX{{\rm B\kern-.05em{\sc i\kern-.025em b}\kern-.08em
		T\kern-.1667em\lower.7ex\hbox{E}\kern-.125emX}}

\begin{document}

\title{Beyond socket options: making the Linux TCP stack truly extensible\\
\thanks{ISBN 978-3-903176-16-4 \textcopyright  2019 IFIP}
}

\author{
	\IEEEauthorblockN{Viet-Hoang Tran}
	\IEEEauthorblockA{\textit{ICTEAM},
		\textit{UCLouvain}\\
		Louvain-la-Neuve, Belgium \\
		hoang.tran@uclouvain.be}
	\and
	\IEEEauthorblockN{Olivier Bonaventure}
	\IEEEauthorblockA{\textit{ICTEAM},
		\textit{UCLouvain}\\
		Louvain-la-Neuve, Belgium \\
		olivier.bonaventure@uclouvain.be}
	}

\maketitle

\begin{abstract}
The Transmission Control Protocol (TCP) is one of the most important protocols in today's Internet. Its specification and implementations have been refined for almost forty years. The Linux TCP stack is one of the most widely used TCP stacks given its utilisation on servers and Android smartphones and tablets. However, TCP and its implementations evolve very slowly. In this paper, we demonstrate how to leverage the eBPF virtual machine that is part of the recent versions of the Linux kernel to make the TCP stack easier to extend. 

We demonstrate a variety of use cases where the eBPF code is injected inside a running kernel to update or tune the TCP implementation. We first implement the TCP User Timeout Option. Then we propose a new option that enables a client to request a server to use a specific congestion control scheme. Our third extension is a TCP option that sets the initial congestion window. We then demonstrate how eBPF code can be used to tune the acknowledgment strategy.
\end{abstract}

\begin{IEEEkeywords}
	TCP, protocol extension, eBPF, dynamic policy
\end{IEEEkeywords}

\section{Introduction}

The Transmission Control Protocol (TCP) \cite{rfc793} remains one of the core protocols in today's Internet. The designers of TCP did not expect that it would be used by billions of devices, but they did foresee the importance of designing an extensible protocol. TCP's extensibility depends on two important factors: $(i)$ the extensibility of the protocol and $(ii)$ the extensibility of its implementations.

To be extensible, the TCP protocol includes TCP options that can be placed in the TCP header. A TCP connection starts with a three-way handshake during which the client proposes a set of extensions as TCP options placed in the SYN packet and the server replies with its supported options. The accepted TCP options can then be attached to the other packets exchanged over this connection. Various TCP extensions have been proposed during the last decades: TCP Timestamp and large windows \cite{rfc1323}, Selective Acknowledgements \cite{rfc2018}, TCP Fast Open \cite{rfc7413}, Multipath TCP \cite{rfc6824} and so on. However, deploying a new TCP option takes time. It needs to be defined, accepted by the IETF and then implemented by major TCP stacks. Measurements show that Selective Acknowledgements took more than a decade to be deployed \cite{fukuda2011analysis} and the Timestamp option is still not enabled by the Microsoft stacks \cite{honda2014rekindling}. More recently, middlebox interference became an important concern \cite{honda2011still}.

The second, and often forgotten, factor is the extensibility of the TCP implementations. For many years, the Unix 4.x BSD stack has served as the reference TCP implementation \cite{wright1995tcp}. When Van Jacobson wrote his seminal paper on congestion avoidance and control \cite{jacobson1988congestion}, his works had a large impact because they were quickly integrated inside this reference implementation. Today, this stack is less popular than the Linux TCP stack that is used by a large fraction of Internet servers and Android smartphones. This Linux stack has been extended to support TCP Fast Open \cite{radhakrishnan2011tcp}, Multipath TCP \cite{raiciu2012hard} and many other TCP extensions. The TCP stack in Linux 1.0 in 1994 contained 3k lines. It grew to 18k lines in version 2.6 (2010). Today's TCP implementation spans more than 80k lines of C code in the Linux kernel. Most of the recent additions to the Linux TCP stack have been driven by the needs of large content providers.

The Linux TCP stack is highly optimised for the most common use cases, but it has very limited ability to \textit{adapt} to a changing environment of network conditions, workloads or user requirements. It can be tuned through a myriad of \texttt{sysctl} parameters \footnote{See \url{https://www.kernel.org/doc/Documentation/networking/ip-sysctl.txt}}. These parameters allow to tune many TCP aspects e.g. delayed ACK timeout, ACKing strategy, congestion control scheme. More importantly, the \texttt{sysctl} interface only allows changing system-wide behaviors, but it does not support per-connection policies. Some of these parameters and others are exposed as socket options\footnote{See \url{http://man7.org/linux/man-pages/man7/socket.7.html}} that can be set by applications on a per-connection basis.

As explained in Section~\ref{related}, some researchers have proposed techniques to extend the Linux TCP stack, but they do not allow to read or write new TCP options. We consider that supporting new TCP options is a crucial part of a truly extensible framework for TCP.
In short, the main contributions of this paper are as follows:
\begin{enumerate}
\item We propose and implement a light eBPF-based framework that enables users to easily add support for new TCP options in the Linux TCP stack
\item We propose four use cases that leverage our framework to adapt the stack to various scenarios or user requirements.
\end{enumerate}

The remainder of this paper is organized as follows: Section~\ref{related} mentions the related work and the objectives of this work. We present our method and implementation of TCP option framework in Section~\ref{method}. Several use cases for new TCP options are presented in Section~\ref{s:tcp-usecase}. 
Section~\ref{discuss} discusses the insights and future work. Finally, Section~\ref{artefact} provides links to the artefacts of our work.

\section{State of the art}
\label{related}

Transport protocols such as TCP can be implemented inside the operating systems' kernel \cite{wright1995tcp} or as a library inside the application. The main motivation of kernel stacks is that a single stack can support all applications, ensure that they do not interfere and achieve high performance \cite{clark1989analysis}. A drawback of in-kernel implementations is that they are more difficult to extend than user space ones. On the other hand, user space implementations are more flexible, but they are often less mature than the in-kernel ones. Recent advances have enabled user space implementations to reach higher performance \cite{mtcp}.

\subsection{In-kernel approaches}

Several researchers have proposed solutions to simplify the extension of in-kernel implementations.  STP \cite{patel2003stp} was an ambitious effort to allow end hosts to load untrusted code from remote peers to upgrade their transport protocols. The conceptual idea of loading user code into a sandbox in the kernel is similar to the utilisation of the eBPF virtual machine in today's Linux kernel. %

The idea of exposing and allowing applications to set internal state variables of TCP connections was early proposed \cite{icTCP,mogul2004}. This permits the control plane of TCP congestion control to be moved from kernel to userland \cite{icTCP,usercc2018}. This also enables adding new non-intrusive features \cite{icTCP}, as long as they do not change the wire format or the internal state of TCP. In terms of performance, this approach requires costly  switching back and forth between userspace and kernelspace for both reading and writing parameters.

\subsection{Userland approaches}

Besides kernel stacks, there are complete user-space TCP stacks \cite{mtcp,lwip,uip}. Their nature makes them be easier to be modified by application developers than the in-kernel TCP stacks. However, they often lack many crucial features (e.g.: PMTU discovery) or the rich ecosystem of cooperative facilities (notably but not limited to \texttt{iptables, namespacing, cgroup}) and debugging utilities. New transport protocols such as QUIC \cite{langley2017quic} were designed with user space implementations in mind. Several QUIC implementations are being actively developed\footnote{See \url{https://github.com/quicwg/base-drafts/wiki/Implementations}.}. The QUIC protocol was designed to be easier to extend than the TCP protocol and its encrypted packets should prevent most types of middlebox interference. Currently, however, a portion of networks block (4.4\%) \cite{langley2017quic} or rate-limit UDP traffic.

The Linux Kernel Library (LKL) \cite{lkl} is a compromise between in-kernel and user space implementations since it wraps a custom Linux network stack into a user library, allowing each application to use a different Linux network stack. This approach allows applications to use new features (e.g.: TCP Fast Open, MPTCP) even if updating the host kernel is not possible or not desirable. However, it currently induces some memory overhead 
and the dynamicity of the network stack was not considered.

\subsection{Linux kernel facilities}

The Linux TCP stack provides the applications several ways to observe or change the state of the underlying TCP connections. This stack provides the \texttt{TCP\_INFO} socket option that returns many state variables. Some mobile applications use it frequently \cite{schueren2017tcpsnitch}. Another example is the fact that congestion controllers are implemented as modules in the Linux TCP stack. The default congestion controller can be set through a \texttt{sysctl} or configured on a per connection basis once loaded.

The Linux kernel includes several facilities which can be used to extend its TCP implementation. First, \texttt{Netlink} \cite{rfc3549} establishes channels between kernel space and user space. It has been used to support user-level control plane for MPTCP path manager \cite{hesmans2015smapp}. However, this approach requires the addition of a lot of code into both kernel and userspace, causing both memory and processing overhead.
 
A low-level way to change the kernel execution path is to use \texttt{kprobes}' capability \cite{kprobes} of changing the register set and instruction pointer. It allows capturing some information when a specific kernel function is executed. However, this approach is highly fragile and prone to error, which could lead to serious consequences such as kernel panics or kernel data leaks.

A comprehensive approach is to build a custom sandbox which allows userspace to load custom code into the kernel and change the behavior of stack, similar to STP \cite{patel2003stp}. %
Classic BPF (cBPF) Virtual Machine has been part of the Linux kernel for more than two decades. It has been mainly used to write filters to capture packets. Recently, this Virtual Machine has been extended and renamed the extended BPF (eBPF). It supports several use cases such as sandboxing system calls (\texttt{seccomp}), tracing kernel events \cite{greg2018linux}, implement hyperupcalls \cite{amit2018design}.
Several networking use cases already leverage eBPF. For example, XDP uses it for fast packet processing \cite{hoiland2018express}, IPv6 Segment Routing uses it to support network programming \cite{xhonneux2018leveraging} and it improves the extensibility of Open vSwitch \cite{tu2017building}. A key benefit of the in-kernel eBPF virtual machine is that each eBPF bytecode is provably verified before being injected inside the kernel. This ensures that eBPF bytecode cannot harm a running kernel.
\section{Extending TCP in the Linux kernel}
\label{method}

As explained earlier, the standard method to extend TCP is to define a new TCP option. In the early days, researchers introduced new TCP options and registered them with the IANA. Then, the IETF took control of most of the evolution of the TCP stack and most recent TCP extensions have been designed with the IETF. Today, researchers willing to deploy a new TCP option cannot anymore simply register their new option within IANA. The IETF has defined a format for experimental TCP options \cite{rfc6994}. This format has not yet been widely used, but we leverage it in this paper to minimize the possibility of middlebox interference when using new TCP extensions. 

\begin{figure} [h]
        \centering
        \includegraphics[width=0.7\linewidth]{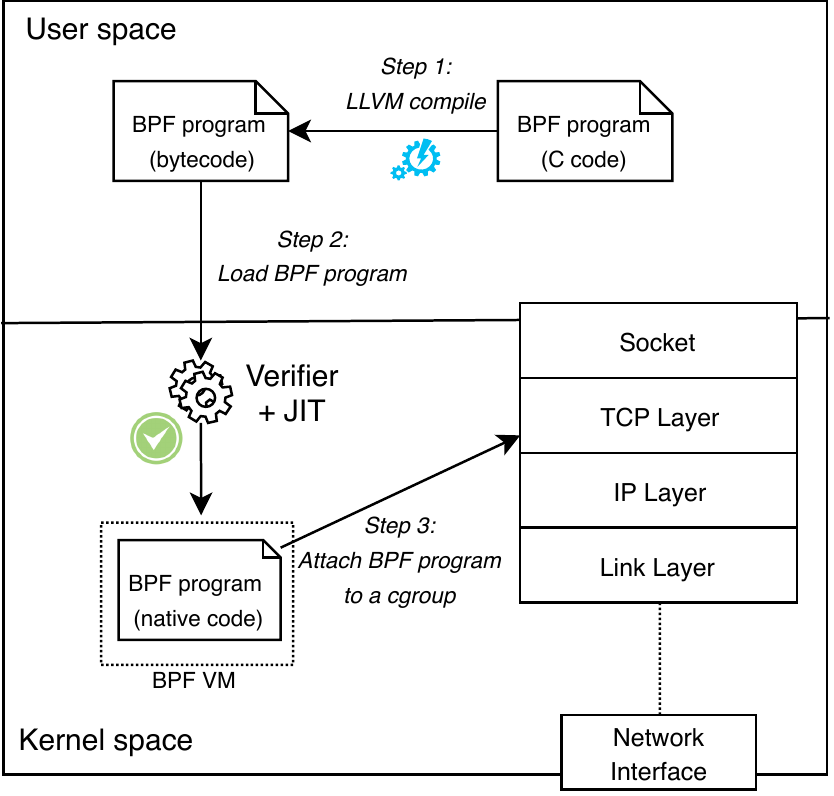}
        \caption{A user application compiles and injects eBPF program into kernel}
        \label{fig:bpf-inject-code}
\end{figure}

From an implementation viewpoint, a TCP extension can be added to the Linux kernel as a set of patches. This approach has been used by many researchers (see e.g. \cite{raiciu2012hard,radhakrishnan2011tcp}). However, users are forced to recompile their kernels with those patches to support the proposed extension. This severely limits their deployment. A better approach is to leverage as much as possible the eBPF virtual machine. Thanks to this virtual machine, any application can inject code inside the underlying TCP stack to modify its behaviour (as shown in Fig.~\ref{fig:bpf-inject-code}). For example, an interactive application running on a smartphone could inject a retransmission technique that is optimised for short packets while a datacenter server could inject another congestion control scheme. This injection could be done directly by the network application or by a system daemon in userspace. Before loaded, the eBPF code needs to be passed through a static verifier to make sure it is both secure and fast. The eBPF code can be executed in an efficient way thanks to the JIT compiling support.

As explained in the previous section, several use cases have been developed for eBPF in the Linux kernel. These address various components of the Linux kernel and many focus on performance monitoring. In 2017, Lawrence Brakmo proposed the TCP-BPF framework \cite{brakmo2017tcp} which is specifically built for the TCP stack and provides basic support to extend the TCP stack. We leverage TCP-BPF as a starting point for our work.

TCP-BPF has been gradually added \cite{tcp-bpf-patch1,tcp-bpf-patch2,tcp-bpf-patch3}
into mainstream kernel in versions 4.13 through 4.15. It was mainly designed to help network administrators to tune the TCP configurations of servers in datacenters at the connection level or forcing the application developers to use specific socket options. The main objective of TCP-BPF was to optimize the TCP parameters in a programmable manner. For example, TCP-BPF would configure the stack to use small buffers and a small SYN retransmission timer for a container that includes applications running inside a given datacenter. However, a different eBPF code would be used for applications that perform bulk transfers between datacenters.

TCP-BPF \cite{brakmo2017tcp} adds several callbacks (also called hooks) to call BPF programs at different stages of a TCP connection. There are two main types of callbacks. The first type is the callbacks at the beginning of each connection: e.g. when the client calls \texttt{connect()} or when the server calls \texttt{listen()} or when the connection is fully established. These callbacks are always enabled. On the contrary, callbacks of the second type are only enabled once they have been requested by a BPF program to limit the overhead when they are unused. These include callbacks triggered when RTO fires, when a packet is retransmitted, or when the TCP state is changed. TCP-BPF allows BPF programs to read and write to many fields of data structures (\texttt{tcp\_sock}) maintained by the TCP stack via a mirror structure \texttt{bpf\_sock\_ops}. It also provides access to other internal TCP variables via indirect \texttt{bpf\_getsockopt()} and \texttt{bpf\_setsockopt()} helper functions. These different hooks call the BPF programs by using the same helper function \texttt{tcp\_call\_bpf()}. Since a BPF program can be called from different places in the kernel, the hooks are also associated with an argument (\texttt{op}) to indicate the callback type to let the BPF program know the current context in the kernel.
        
Since TCP-BPF was implemented by Facebook engineers to work in data center environment, it requires \texttt{cgroup} version 2 to manage various system resources such as CPU or memory for their containers.
For this reason, it is necessary to attach and the BPF program to the same \texttt{cgroup-v2} of the user application. However, this is not a permanent requirement, rather it should be considered as an implementation caveat and can be changed later.

\subsection{Supporting a new TCP option}
\label{s:support-tcp-opt}

As an illustration of how it is possible to use eBPF programs to extend the Linux TCP stack, we first describe the changes that are required to add the support for a new TCP option. Table~\ref{hook-list} summarizes new hooks added by our framework and their meaning.

\begin{figure} [t]
        \centering
        \includegraphics[width=\linewidth]{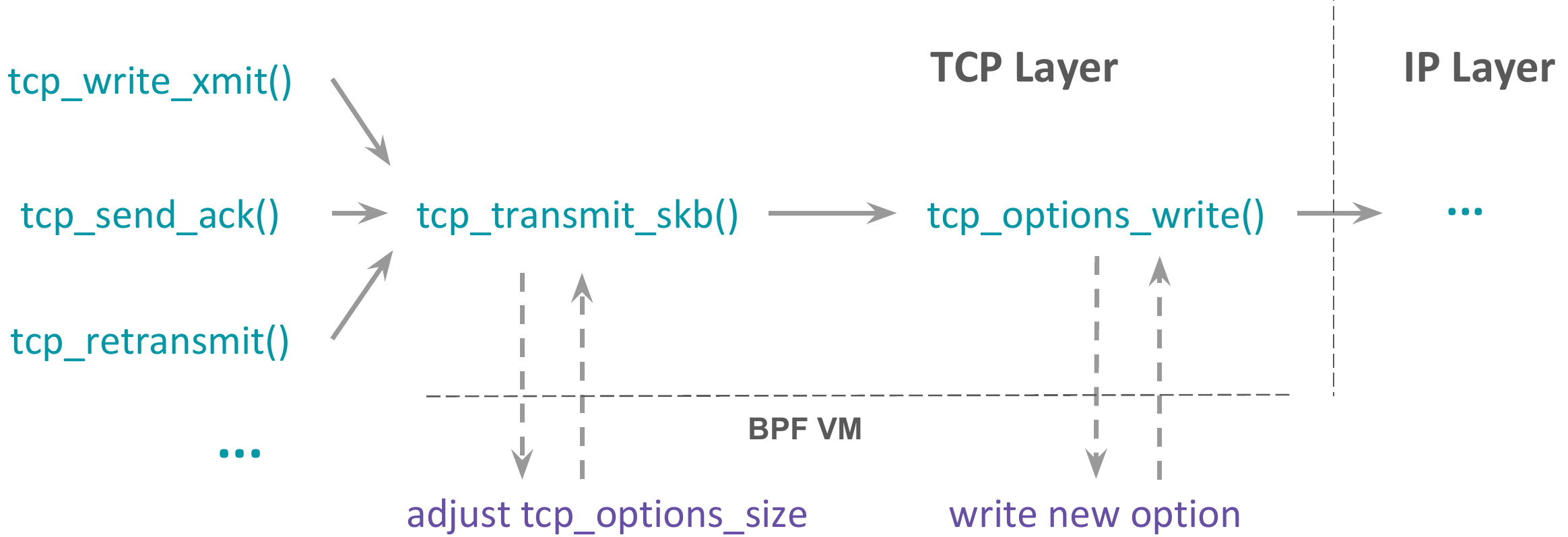}
        \caption{Insert TCP options to outgoing packets}
        \label{fig:insert-tcp-options}
\end{figure}

Let us first analyse the sender side. When sending packets, the \texttt{tcp\_transmit\_skb()} function in \texttt{tcp\_output.c} creates the TCP header and the required TCP options. TCP options are written in two steps: $(i)$ the stack computes the size of all provisioned TCP options and $(ii)$ it writes the TCP options in \texttt{tcp\_options\_write()}. Therefore, to insert a new TCP option we add two separate hooks into above places, as illustrated in Fig.~\ref{fig:insert-tcp-options}.

We first add in \texttt{tcp\_transmit\_skb()} a hook which calls a BPF program to adjust the provisioned size of all TCP options (\texttt{tcp\_options\_size}). We also verify that it does not exceed 40 bytes - the maximum option size. Then, at the end of \texttt{tcp\_options\_write()}, a second hook calls a BPF program which passes the new option data to the kernel. The kernel is then responsible for writing the new option data at the current option pointer.

These hooks are only activated when the BPF program sets the appropriate flag (per connection in struct \texttt{tcp\_sock}, as explained below).

There is still one thing the framework has to take care of. Since the TCP stack calculates the current MSS at multiple places, the composed packets may be too large and could be fragmented on the wire. We need to update \texttt{tcp\_current\_mss()} function to take the length of to-be-added option into the consideration. This is performed by a hook with the same \texttt{op} type as the above hook (which adjusts \texttt{tcp\_options\_size}) that is added to \texttt{tcp\_current\_mss()} and thus is completely transparent to the BPF programs.

\begin{figure} [t]
        \centering
        \includegraphics[width=0.85\linewidth]{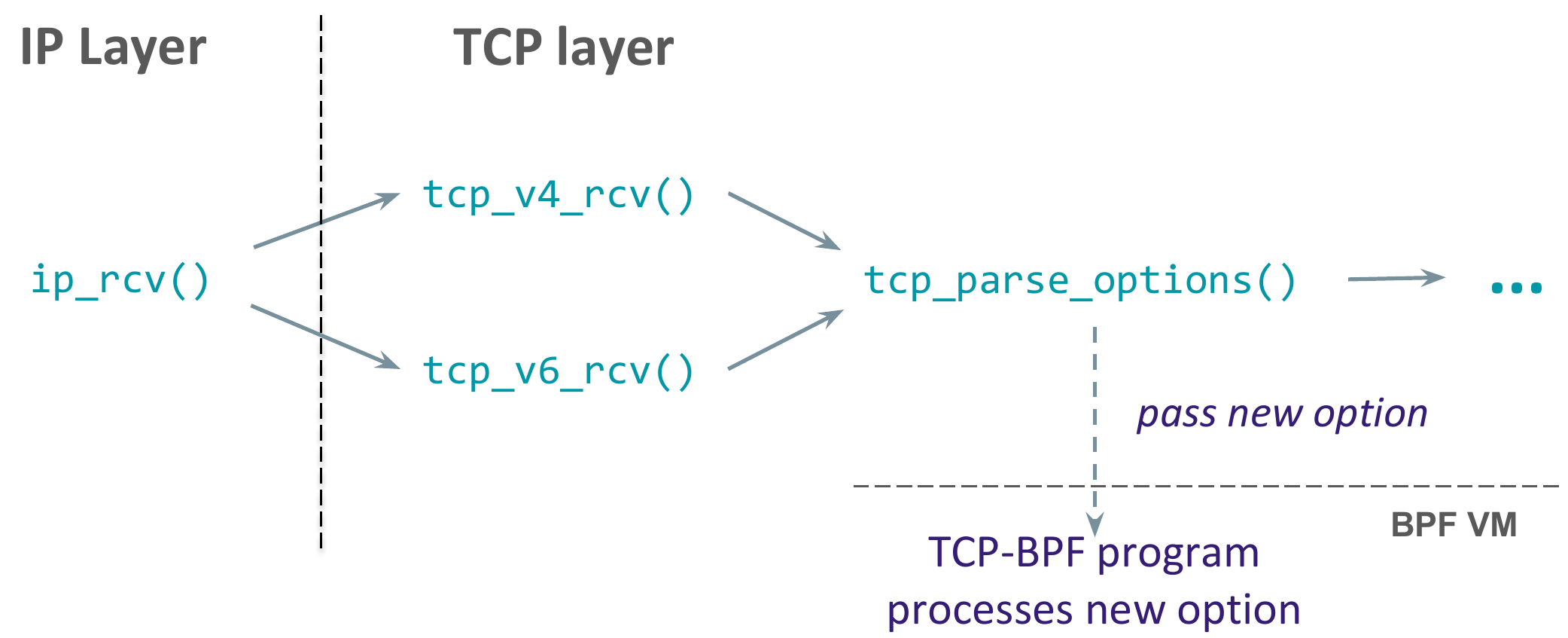}
        \caption{Pass unknown TCP options of incoming packets to BPF program}
        \label{fig:parse-tcp-options}
\end{figure}

On the receiver side, the extension is simpler. Linux TCP parses the options of incoming TCP packets in \texttt{tcp\_parse\_options()}, in which all unknown options are ignored.   At the end of this function, we added a hook to pass these unknown options to the BPF program, as shown in Fig.~\ref{fig:parse-tcp-options}. This hook, once activated, will pass the option data along with option kind and length to the BPF program. The hook could also pass multiple new options of the same TCP packet to one or multiple BPF programs. The BPF program reads the option and applies a relevant change to the TCP socket, e.g. by setting socket values via \texttt{bpf\_sock\_ops} or \texttt{bpf\_setsockopt()}.

By building on top of TCP-BPF, we can implement our framework with modest changes to the kernel (75 LoCs). TCP option insertion support requires around 60 LoCs, while the TCP parsing support requires only 15 LoCs since it is much simpler as explained above.
Table~\ref{Loc} lists the size of our framework and each use case with regards to the number of lines of code (LoC) changed in the kernel.

We added a minor kernel change to support getting and setting internal TCP user timeout value directly in eBPF program, while current kernel has already supported setting and getting Congestion Control algorithm or Initial Window. The implementation to support configurable TCP Delayed ACK, which is essentially based on an RFC patch \cite{delayed-ack-param}, is reasonably larger.

\begin{table} [h]
\footnotesize
\begin{tabular}{ l r r }
    \toprule
    & \textbf{Kernel changes} & \textbf{BPF program} \\
    \midrule
    TCP Option framework           &             75 &           - \\
    Use case: TCP User Timeout     &             16 &          76 \\
    Use case: Congestion Control   &              0 &          92 \\
    Use case: Initial Window       &              0 &          76 \\
    Use case: Delayed ACK          &             94 &          77 \\
    \midrule
\end{tabular}
\captionsetup{justification=centering,margin=0.7cm}
\caption{Lines of code (LoC) of the framework and each use case}
\label{Loc}
\end{table}

\bgroup
\def\arraystretch{1.1}

\begin{table*}
\scriptsize
\centering
\begin{tabular}{l l l l} \hline
\textbf{Hook}                          & \textbf{In kernel function}        & \textbf{Passed arguments}          & \textbf{Meaning}                                  \\ \hline
\texttt{BPF\_TCP\_OPTIONS\_SIZE\_CALC} & \texttt{tcp\_transmit\_skb}        & Size of all TCP options            & Let BPF program to get and adjust                 \\
                                       & \texttt{tcp\_current\_mss}         &                                    & the length of all TCP options in a packet         \\ \hline
\texttt{BPF\_TCP\_OPTIONS\_WRITE}      & \texttt{tcp\_options\_write}       & -                                  & Let BPF program to insert new TCP option          \\ \hline
\texttt{BPF\_TCP\_PARSE\_OPTIONS}      & \texttt{tcp\_parse\_options}       & Kind, size, and data of new option & Pass unknown TCP option to BPF program            \\ \hline
\end{tabular}
\caption{New BPF hooks added by TCP option framework}
\label{hook-list}
\end{table*}

\egroup

\subsection{How to select the desired packets for inserting new option?}

The first question is how to select the relevant connections. A user daemon can specify the cgroup that the targeted connections are associated with, before loading the BPF program. At runtime, the BPF program can check the 4-tuple of IP addresses and ports to only take care of the interesting connections. These operations have already been supported by the vanilla kernel so no kernel change is required.

The second question is how to insert new options to the desired packets only. To mark when the program wants to actually insert new options, we need to add a new flag. TCP-BPF already has a flag array (\texttt{bpf\_sock\_ops\_cb\_flags}) in the \texttt{tcp\_sock} struct, for enabling and disabling the hooks at different phases of a TCP connection. We extend this flag array with our flag to minimize the amount of changed code. The BPF program can set the flag at one hook (e.g. when the connection is fully established) to enable option writing onto all following \texttt{skb}s of the same TCP connection, and unset the flag at another hook (e.g.: when RTO fires) to disable option writing from this point.

\subsection{Performance Overhead}

Linux TCP is a high-performance stack. Any proposed extension should take the performance impact into consideration. To evaluate the performance impact of our BPF extensions, we run the iPerf3 \cite{iperf3} test between two dedicate machines over a 10~Gbps link. Each machine is equipped with an Intel Xeon X3440 2.53GHz CPU and 16~GB RAM. Our framework is implemented in Linux kernel version 4.17-rc5. We use different TCP-BPF programs that are called to manipulate \emph{each} transmitted packet. We consider four different experiments.

\begin{enumerate}
    \item Baseline, no BPF program is loaded
    \item A BPF program inserts a new TCP option on the sender
    \item A BPF program on the sender (to insert a new option) and one on the receiver (to parse this new option)
    \item A BPF program on the sender that inserts a new option while the receiver parses this option and then calls both \texttt{bpf\_setsockopt()} and \texttt{bpf\_getsockopt()}
\end{enumerate}

\begin{figure*}
    \centering
    \begin{subfigure}{0.66\columnwidth}
        \centering
        \includegraphics[width=1.05\linewidth]{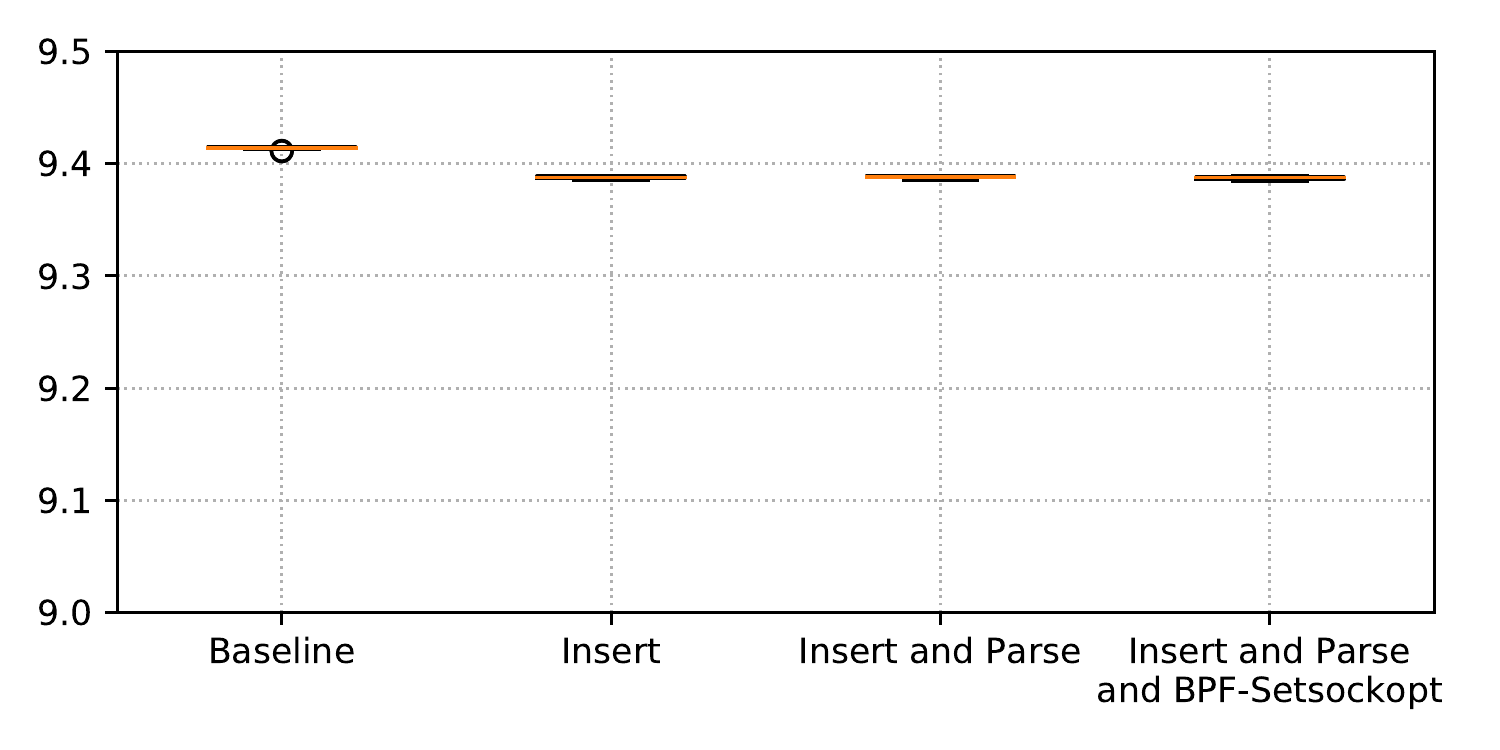}
        \caption{Average Throughput (Gbps)}
        \label{fig:overhead-tput}
    \end{subfigure}
    \hfill
    \begin{subfigure}{0.66\columnwidth}
        \includegraphics[width=1.05\linewidth]{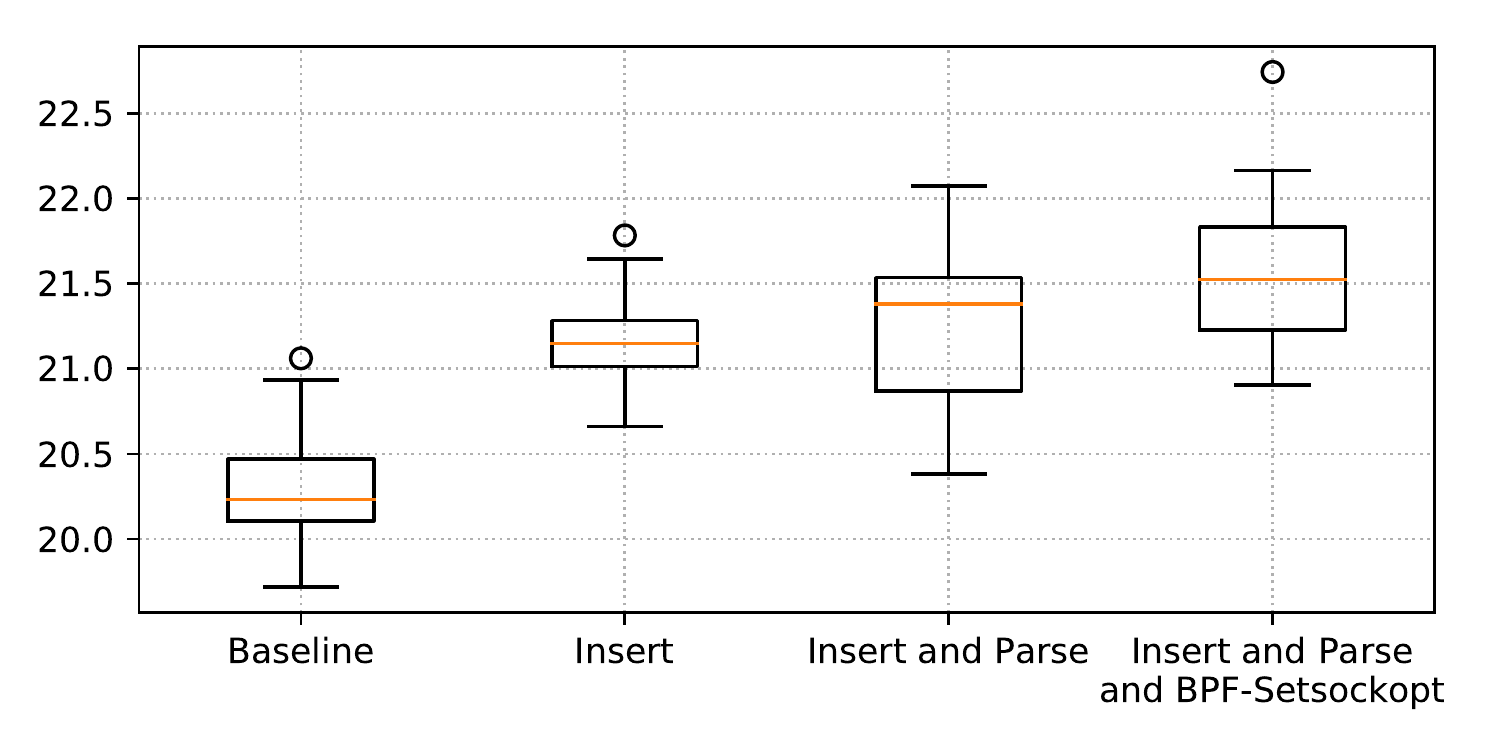}
        \centering
        \caption{Sender's CPU Usage  (\%)}
        \label{fig:overhead-local-cpu}
    \end{subfigure}
    \hfill
    \begin{subfigure}{0.66\columnwidth}
        \includegraphics[width=1.05\linewidth]{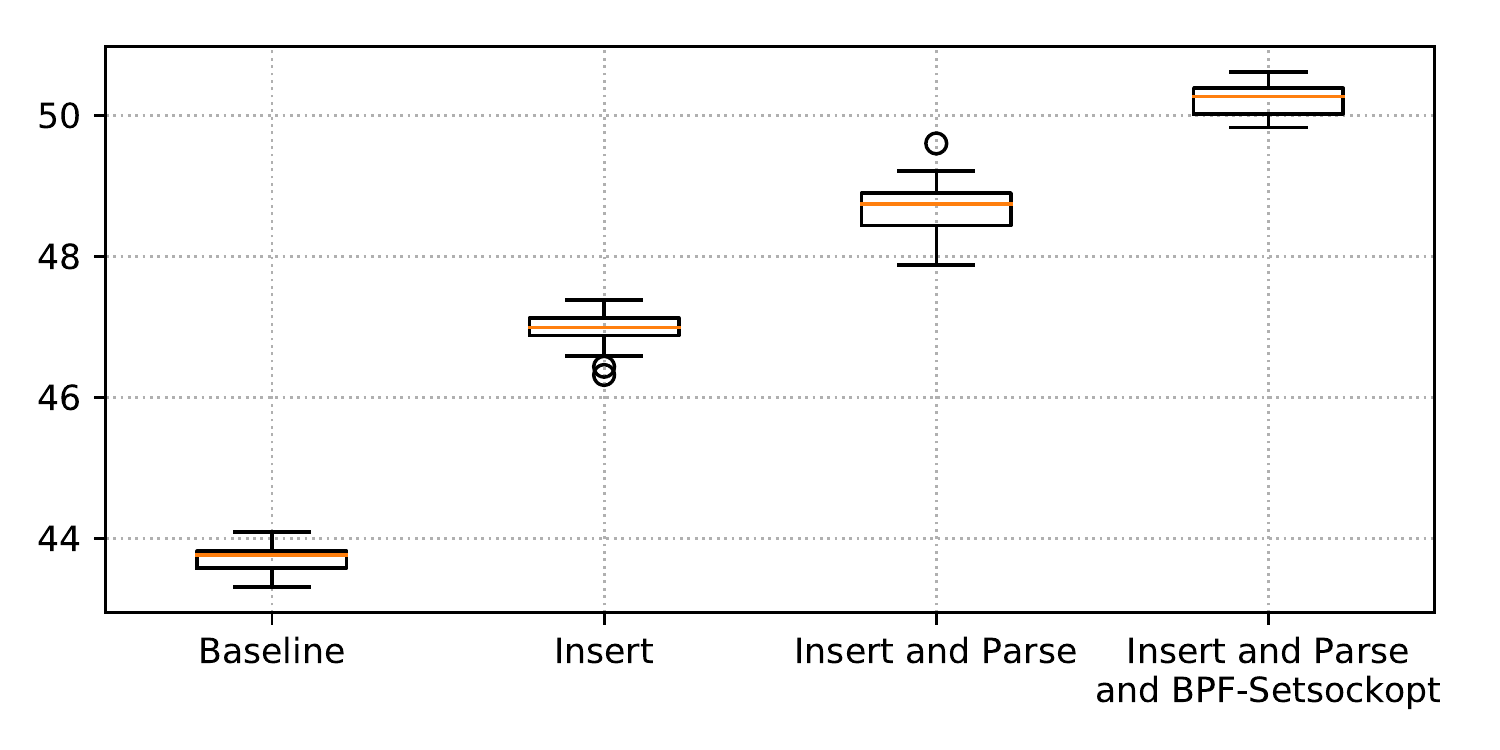}
        \centering
        \caption{Receiver's CPU Usage (\%)}
        \label{fig:overhead-remote-cpu}
    \end{subfigure}
    \captionsetup{justification=centering}
    \caption{Benchmarking results: iPerf3 client and server test over 10Gbps link}
    \label{overhead}
\end{figure*}

Each measurement lasts 40 seconds and each scenario is repeated 20 times. Figure~\ref{overhead} shows the benchmark results reported by iPerf3 for each situation. The average throughput is reduced from 9.41~Gbps in the baseline case to 9.38~Gbps in all three BPF-enabled scenarios, mostly because our newly inserted TCP option has increased the TCP header size.
Meanwhile, there is no statistically meaningful difference of round-trip-time among all cases (all around 410~microseconds) therefore we do not present them here. The CPU utilisation overhead is the most noticeable one which is about 10\% in the worst case, as shown in Fig.~\ref{fig:overhead-local-cpu} and Fig.~\ref{fig:overhead-remote-cpu}.

To push the TCP stack to the limit, we conducted another extreme benchmark with the iPerf3 client and server on the same host machine. This benchmark tries to send as much data as possible via the loopback interface to saturate the TCP stack, which is an extreme but unrealistic scenario. We do not show its figures here due to lack of space. The average throughput obtained with baseline tests is 30.1~Gbps (about 2.5~Mpps) and the average RTT is 27.1~usecs. Using a BPF program that inserts a new TCP option introduces a throughput reduction of about 12.7\% and a delay increment of 14.8\% (4~usecs). Using a BPF program that parses a new TCP option reduces further the throughput by 3.8\% and increases the delay by 4.5\%. Calling operations such as \texttt{bpf\_getsockopt} or \texttt{bpf\_getsockopt} does not have a noticeable impact.

These results suggest that most of the overhead of the framework comes from the call-backs to the BPF program, not from the execution of the BPF program itself.
\section{TCP Use Cases}
\label{s:tcp-usecase}

In this section, we illustrate with several different use cases how it is possible to leverage BPF programs to extend the Linux TCP stack. We start in Section~\ref{ss:uto} with the TCP User Timeout \cite{rfc5482} that has not been completely implemented in the Linux TCP stack. We then propose and implement in Section~\ref{ss:cong} a TCP option that enables a client to suggest the congestion control scheme to be used by a server. We then propose a TCP option to set the initial congestion window in Section~\ref{ss:iw} and finally one to request a specific acknowledgement strategy in Section~\ref{ss:delack}.

\subsection{Use case: TCP User Timeout Option}
\label{ss:uto}

The TCP User Timeout (UTO) option \cite{rfc5482} was proposed to allow a host to inform its peer of the maximum time that data could remain unacknowledged before forcing the termination of the associated connection. There are two typical use cases for this option. First, an application that wants to survive transient failures would select a very large UTO. Second, a mobile interactive application that is used on smartphones equipped with Wi-Fi and cellular interfaces could use a short UTO (e.g. one second) to quickly detect connectivity problems and switch to the other network interface. %

The UTO option \cite{rfc5482} carries the suggested timeout value. It is sent unreliably, typically inside a TCP ACK. In contrast with most TCP extensions, the utilisation of this option is not negotiated during the three-way handshake. It is simply used once the connection has been established. Linux allows applications to set the maximum value of the retransmission timers through the \texttt{SO\_RCVTIMEO} and \texttt{SO\_SNDTIMEO} socket options. However, it does not support the UTO option. In Linux, when the UTO timer fires, the kernel signals a timeout error to the user application and changes the connection state to \texttt{TCP\_CLOSE}. However, it is the responsibility of applications to do RST the connection.

On the client side, we implement the UTO option support with a BPF program (76 lines of C code) using our option-writing hooks described in the previous section. On the server side, when it receives a UTO option from the peer, the kernel stack passes the option to a BPF program that parses the option and sets the local socket timer value by leveraging the \texttt{bpf\_setsockopt()} helper function. We also extend \texttt{bpf\_getsockopt()} helper function to query the current User Timeout value of the connection.

\subsection{Use case: TCP Congestion Control Option}
\label{ss:cong}

The Linux TCP stack supports a dozen of pluggable congestion control modules \cite{corbet2005pluggable}. Depending on its configuration, a Linux host may directly support two to three TCP congestion control schemes, e.g. NewReno \cite{rfc5681}, CUBIC \cite{ha2008cubic}, or Vegas \cite{brakmo1995tcp} or BBR \cite{cardwell2017bbr}. Content Distribution Networks (CDN) often tune their congestion control scheme to better serve their customers \cite{bbr_spotify}. However, a given CDN supports a variety of customers and a congestion control scheme that works well to serve a user connected through an optical fiber might not work well for a user connected over a slow ADSL link. Some CDNs tune their TCP stack on a per-prefix basis, but there are many situations where the client that downloads information from a server has much better knowledge of the performance of its access network than the server. For example, a smartphone can easily collect statistics about the amount of reordering and the delay variations that it has observed recently. Based on this information, it could suggest a specific congestion control scheme to be used by a given server.

In our implementation, each supported TCP congestion control scheme is identified by an integer. The mappings between the TCP congestion control schemes and their identifiers could be distributed together with the Linux kernel. 

Our BPF programs on both the client and the server store the list of congestion control algorithms in an array map. This map contains algorithm IDs as the keys and the string names as the corresponding values. When the server receives the congestion control option, the BPF program extracts the identifier and looks it up in the map to retrieve the name of the requested algorithm. It then changes the congestion control scheme applied to this connection using the \texttt{bpf\_setsockopt()} helper function.

To illustrate the utilisation of this congestion control option, we set up the emulation environment similar\footnote{We do not use Mininet directly but use directly built-in facilities in Linux (\texttt{netns}, \texttt{tc},...) because Mininet uses \texttt{cgroup v1} while \texttt{cgroup v2} is currently required by \texttt{tcp-bpf} framework.} to Mininet \cite{lantz2010network}. We set up separate network namespaces for client and server, a Linux bridge in-between, and using Traffic Control (TC) with HTB qdisc to set link bandwidth to 8~Mbps and 40~ms delay per direction. Our emulated client downloads the same large file using the \texttt{curl} software. We use our BPF program to insert in the third ACK packet the TCP congestion control option to request the utilisation of a specific congestion control scheme by the server.

\begin{figure}[tbph!]
	\centering
	\includegraphics[width=0.75\linewidth]{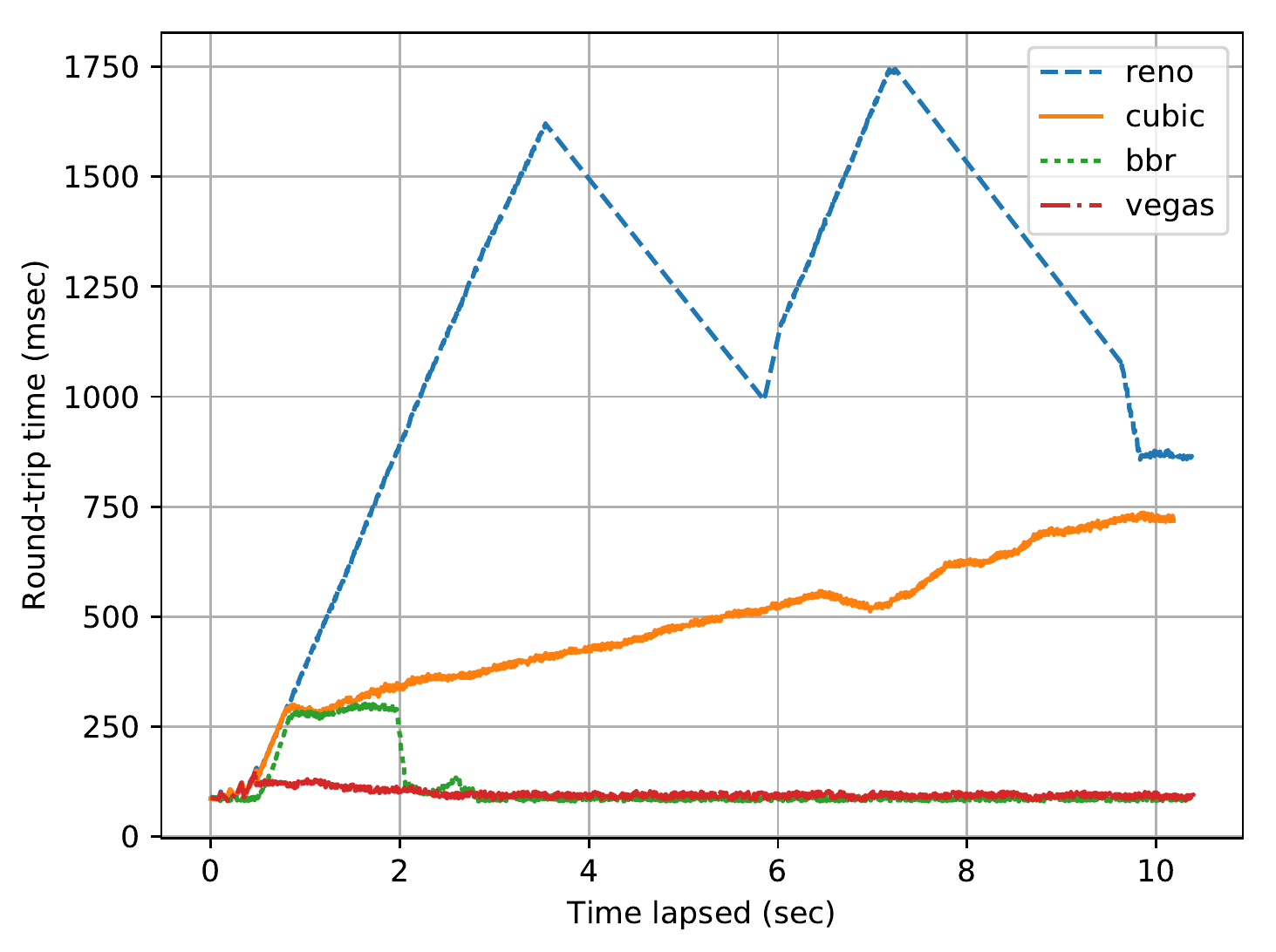}
	\captionsetup{justification=centering}
	\caption{Congestion Control Option test: RTT on the server
		(8~Mbps bandwidth, 40~ms link delay)}
	\label{fig:server-rtt-cc}
\end{figure}

We consider NewReno \cite{rfc5681}, CUBIC \cite{ha2008cubic}, Vegas \cite{brakmo1995tcp} and BBR \cite{cardwell2017bbr} in our experiments. These four congestion control algorithms correctly use the 8~Mbps link, but they differ in the amount of bufferbloat that they cause. Figure~\ref{fig:server-rtt-cc} plots the round-trip-times measured by the server for each congestion control scheme. We repeated the tests multiple times, but they produced nearly identical graphs. Vegas and BBR, the delay-based algorithms, have the lowest Round-trip times (RTT) which are close to the two-way link delays. While Cubic escaped the slow-start phase early, it does not stop RTT from increasing. Among all, NewReno performs worse in terms of delay.

In this example, we used the congestion control option to exchange the identifier of congestion control scheme that the peer should use. The same option could also be extended to provide some parameters of the congestion control scheme. For example, Google QUIC \cite{langley2017quic} uses a variant of CUBIC that is more aggressive than the standard one. This was motivated by the fact that a QUIC session is equivalent to several HTTP/1.1 sessions since it supports streams. The same applies to HTTP/2 running over TCP. 

\subsection{Use case: Option to Request Initial Congestion Window}
\label{ss:iw}

While the congestion control algorithm has a significant impact on the performance of long flows, the selection of Initial congestion window (IW) decisively affects the flow completion time for short flows. This clearly applies to web traffic. The standard IW value has been increased over the years from 2 MSS to 4 MSS \cite{iw4} and later 10 MSS\cite{iw10,iw10-rfc} to keep up with typical network speeds without harming the robustness of the whole system. However, a fixed value cannot adapt to various network conditions. On long fat networks, the sender usually takes a lot of time to reach the congestion avoidance state. But the same IW value may be too large in highly congested networks.

Recent large-scale measurements \cite{iw-imc2017,iw-tma2018} show that while most web servers use the default values of their TCP stacks, CDN operators usually apply much larger values of IW \cite{iw-tma2018}. These researches also suggested that some CDNs customize their IW configuration based on the network type and/or the content type.

Brakmo suggested \cite{brakmo2017tcp} to heuristically select the IW based on the IP prefix using TCP-BPF, with a simple example \cite{iw-bpf}. We extend this approach by defining a new TCP option that lets a client specify its desired IW value. In many deployments, the receivers have more information about the impact of the IW than the senders by observing packet losses at the beginning of connections. However, this opens up the possibility that the malicious peers may use this option to leverage DoS attacks. To deal with this class of attack, we use two mitigations. First, we restrict that this option can be sent only in the SYN-ACK or third ACK of the three-way handshake, but not in the first SYN packet. This also helps implementing the server side more easily since the Linux TCP initializes the full socket only after the completion of the 3-way handshake. Second, the sender needs to verify the peer is from a trusted IP prefix before setting the requested IW value. This client IP verification could be done directly in the BPF program.
The BPF program can also combine client requests with local policies, e.g. take the content type into account when selecting proper IW for the connection. 

To demonstrate the impact of tuning the initial congestion window with web traffic, we use the methodology proposed by Wang et al. \cite{howspdy} with the \texttt{epload} software \cite{epload}. This enables us to emulate real web contents and gather web page download times. 

We set up a similar testbed to the one of the previous use case in Section~\ref{ss:cong}. The path between client and server was configured with 40 Mbps of bandwidth and 40 msec of delay per direction. The server uses \texttt{nginx} to serve the mirrored web contents of top Alexa 170 websites list. On the client side, we ran the \texttt{epload} tool that analyses the dependency graph of web objects, which were recorded with the Chrome browser console, and replays fetching web resources. Every test with each website is repeated three times.

\begin{figure}
	\centering
	\includegraphics[width=0.75\linewidth]{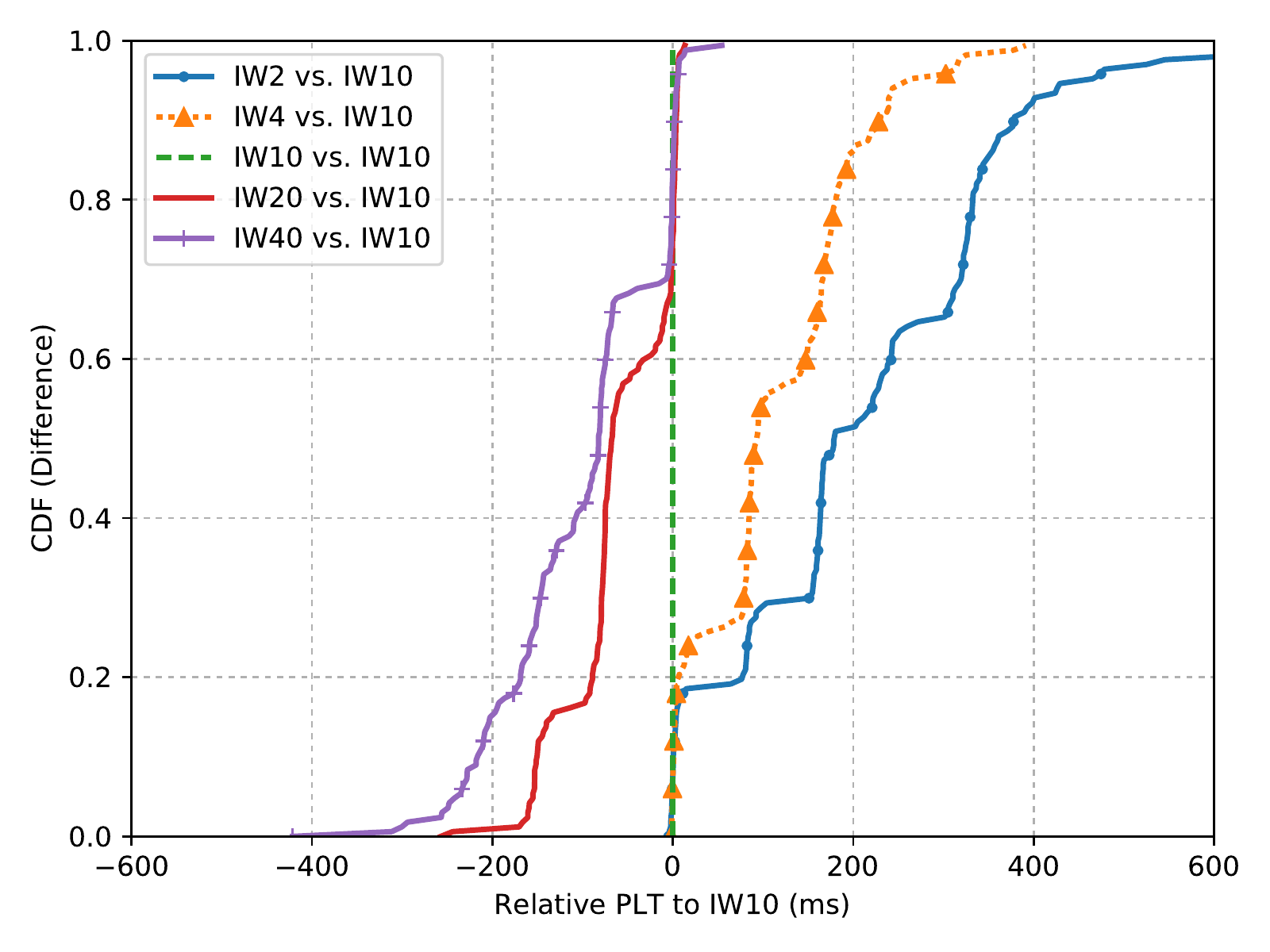}
	\captionsetup{justification=centering}
	\caption{Initial Window Option test: Page Load Time \\
		 relatively to IW=10 (40~Mbps bandwidth, 40~ms link delay)}
	\label{fig:iwspeedup}
\end{figure}

Figure \ref{fig:iwspeedup} shows the relative Page Load Time (PLT) results of each IW value, which is the difference of the Page Load Time between the tests with tuned IW value and the tests with the default IW value (10 MSS) for each website. For about 70\% of websites, the increase of IW yields better Page Load Time, and a few of sites suffered from a higher value of IW, notably when IW is 40. On complex pages that comprise hundreds or thousands of web objects, large IW may cause the link saturated and congested, therefore the PLT is increased. With high network capacity in the experiment, we did not observe much congestion; however, the results could change if the network resource is more limited. Therefore, these results do not suggest that increasing IW always produces better performance, but to show how flexible the Linux TCP stack can be.

\subsection{Use case: Tuning the acknowledgement strategy}
\label{ss:delack}

As a reliable protocol, TCP crucially relies on the ACK packets to detect losses and control the data transfer. Sending ACKs too frequently may impose too much overhead in wireless networks or on fat pipes. On heavily loaded servers, the ACK processing may consume as much as 20\% of the CPU cycles \cite{chan2013tcpack}. On the other hand, sending too few ACKs could probably harm the performance of traditional congestion controls like Reno/Cubic: slow down the increase of congestion window in the slow-start phase, trigger bursty transmissions, overestimate RTT and RTO, or prevent Fast/Early Retransmit recovers from real losses.

For these reasons, the IETF in RFC2525 (section 2.13) \cite{rfc2525} recommended a trade-off: do not delay ACK for more than 500 ms and immediately send ACK for every second packet. Linux follows this recommendation and has hard-coded the minimum and maximum values of the delayed ACK timeout at 40 ms and 200 ms.

However, such fixed values cannot adapt to connections which have very different delay, bandwidth and loss characteristics. They may be too large for local connections, but too small for inter-continental connections. The only customization supported by Linux is to disable the delayed ACK mechanism for each route \cite{per-route-quickack}. However, there is no way for a sender to know the acknowledgement strategy that is used by its peer.

In low-latency environments, the delayed acknowledgement timer causes too many spurious retransmission timeouts, harming the performance. The measured RTTs are inflated by the delayed ACK timeout. The RTO calculation is based on sRTT, so RTO may also be over-estimated by delayed ACKs. There are two separate reasons for this: $(1)$ the default delayed ACK timeout is set too high, and $(2)$ the sender has no information about the delayed ACK behavior on the receiver. For example, in datacenters, the typical RTT is in the order of a few milliseconds, so the estimated RTO is likely dominated by the delayed ACK timeout which is 40 ms at minimum in Linux. While Linux tried to guess delayed ACK to exclude from RTT sampling, there is no reliable way to do this.

Meanwhile, modern networking stacks have adapted to the stretch ACK technique. First, popular networking stacks support pacing, which helps to avoid the bursty transmission issue, a side-effect of the interaction between the stretched ACKing and the classical congestion controls. Second, the congestion control implementations were adapted to increase the congestion window properly with stretch ACKs \cite{stretchACKslowstart,stretchACKpatch}. Furthermore, the Recent ACK (RACK) \cite{rack} (subsumed Tail-Loss Probe (TLP) \cite{dukkipati2013tail}) mechanism which is being standardized and deployed in Linux and Windows \cite{WindowsTCP}. This allows TCP senders to detect losses quickly based on a per-packet timer instead of using duplicated ACKs, reducing the impact of stretch ACK.

Google proposed a TCP Option \cite{googleLowLatencyOption} to negotiate a custom delayed ACK timeout during the three-way handshake. However, as discussed in IETF99 TCPM WG meeting\cite{david_discuss}, there are several issues with this proposal: $(1)$ it is an absolute value, which must be defined before the establishment of the connection, so it cannot adapt to different environments.  Even a well-thought heuristic cannot match all network conditions. $(2)$ A malicious middlebox on the path could inject weird values to drive the hosts into abnormal states. $(3)$ The negotiation uses the SYN and SYN-ACK packets, which may have not enough TCP option space.

We define a similar TCP Option, but with different semantics. Our option contains two fields: $(i)$ the delayed ACK value as a fraction of the minimum RTT and $(ii)$ the amount of unacknowledged data (in units of MSS) that should trigger an immediate ACK. To allow the sender to properly adjust its congestion window during the slow-start, out-of-order receive or retransmission phases, we still keep the original Linux acknowledgement strategy during these phases.

eBPF helps us to change the strategy or parameters dynamically based on the current situation, for example, a client on a crowded wireless network or a server that is sending heavily.

\section{Discussion}
\label{discuss}

TCP was designed to be extensible by using TCP options. However, the last decades have shown that it remains very difficult to extend TCP by defining such a new option. While the IETF has reserved a set of option types for experimental options \cite{rfc6994} to avoid the middlebox interference, TCP implementations such as the Linux TCP stack are monolithic and difficult to extend. In this paper, we have leveraged the eBPF virtual machine in the Linux kernel to demonstrate that it becomes possible to incrementally extend the Linux TCP stack. %
Our work has shown that, with little changes to the kernel code, it is possible to leverage eBPF programs to quickly implement a range of new TCP features.
The main drawback of this method is the limitation of TCP option space, which cannot be larger than 40 bytes. An ongoing effort to overcome this limitation is TCP Extended Data Offset Option\cite{touchEDO}.
On the other hand, it should be considered as a first step to make the Linux TCP stack truly extensible.

The results described in this paper open different directions for future work. A first direction is improving the eBPF support in the Linux kernel. Our implementation is based on the TCP-BPF framework which currently relies on  \texttt{cgroup} version 2. It could be interesting to remove this restriction. More generally, eBPF-based solutions are currently limited by technical constraints which are imposed to guarantee the performance and responsiveness of the kernel. A BPF program cannot contain more than 4096 instructions, BPF functions cannot have more than 5 arguments, loops are not allowed and there are no built-in data structures to implement queues or stacks, even in the latest Linux releases. These restrictions make it difficult to implement complex features, forcing the utilization of workarounds such as using multiple BPF programs that linked together by BPF tail calls. Most of them are not architectural or design flaws but current technical caveats. A map-based implementation of queues and stacks has been merged in net-next branch and should be available in Linux kernel version 4.20. The BPF subsystem maintainers and contributors are working hard to support constrained loops \cite{bpf_loop}, and to extend the program size limit to one million instructions.

A second direction is to actually use eBPF to extend TCP in real deployments. On the public Internet, adding new TCP options remains difficult given the prevalence of middleboxes \cite{honda2011still}. However, TCP is also widely used inside enterprise networks, datacenters and in controlled environments where there is no middlebox interference. It is also used between proxies such as Hybrid Access Networks \cite{mptcp2016deploy} or between edge servers and core servers of CDNs. Furthermore, there is anecdotal evidence that large content providers use a tuned version of the Linux TCP stack that has diverged from the mainline Linux kernel over the years. This implies that either they frequently need to backport new features of the Linux kernel or do not use these improvements in their stack. Using eBPF would enable them to both completely tune their Linux TCP stack and still benefit from the community improvements. 

A third and more interesting direction in the long term would be to make the Linux TCP stack completely modular. It currently contains a wide range of heuristics and optimisations such as congestion control, retransmission techniques, loss detection heuristics, automatic buffer tuning. All these heuristics could be implemented as eBPF programs to enable applications to replace or tune them based on their requirements.

Finally, this approach could be applied to extend other protocols that support an optional field e.g. IP option fields, UDP option fields, or SCTP chunk headers. However, they are also susceptible to the middlebox interference and the implementers and the users need to take it into consideration.

\section{Artefacts}
\label{artefact}
The implementation of our TCP option extension framework, different use cases and the experiment scripts is publicly available at \url{https://github.com/hoang-tranviet/tcp-options-bpf} and \url{https://github.com/hoang-tranviet/Epload}.
Our analysis and plot scripts is available at \url{https://github.com/hoang-tranviet/tcp-options-bpf-analysis}.
Experiment results and related resources can be found at \url{https://www.info.ucl.ac.be/~tranviet/}.

\section*{Acknowledgements}

This work was supported by the ARC-SDN project and the Wallinnov MQUIC
project. We thank Olivier Tilmans for giving us useful explanations and suggestions.

\bibliographystyle{plain}
\bibliography{paper-bpf,rfc}

\end{document}